# Secured Position Location and Tracking (SPL&T) for Detection of Multiple Malicious Nodes Maintaining Two Friendly References in Mobile Ad hoc Networks

**Niraj Shakhakarmi, Dhadesugoor R. Vaman**

**Department of Electrical & Computer Engineering, Prairie View A&M University (Texas A&M University System)**
**Prairie View, Houston, Texas, 77446, USA**

**Abstract**

Secured Position Location and Tracking (PL&T) scheme is developed for multiple malicious radios or nodes detection using integrated key based strict friendly scheme and position location and tracking by multi-sectored based multiple target's PL&T. The friendly and malicious nodes detection is based on the integrated key consisting of symmetric keys, geographic location and round trip response time. Two strictly friend references dynamically form the tracking zone over the detected multiple malicious nodes using the multi-sectored adaptive beam forming. This PL&T technique is robust, precise, scalable, and faster than using the single reference, two reference and three reference nodes based PL&T method in the battlefield oriented Mobile Ad hoc Networks. The simulation results show that the lower relative speed bound of any participating node increased the switching overhead, the decreasing received energy with increasing number of the multi-sectored beams reduced tracking accuracy and the strict friendly authentication overhead depends upon the time period between two latest periodic authentication failures.

## 1. Introduction

The future generation of the digitized battlefield MANET and sensor networks provide robust, high-speed, contention free network architecture and overcome the line of sight limitation as well as jamming threats. It also needs to do secured Position Location and Tracking (PL&T) of the friendly radios or nodes protecting from the hostile nodes, and preventing the hostile nodes from accessing the secured end-to-end paths. This is executed by maintaining secured multi-hop connection with strict friendly nodes which deploy the integrated approach of friendly authentication, the use of novel PL&T method and encryption of PL&T information. The PL&T of hostile node is also computed and multi-casted to friendly nodes, till the hostile node participated in PL&T.

The proposed secured PL&T approach includes the friendly and malicious nodes detection by the strict friendly verification and the location tracking by zone forming, adaptive beam forming and triangulation. The multiple malicious targets' PL&T is developed using the multiple adaptive beam forming over the rapidly predicted multiple zones which dynamically allow tracking for random trajectory as well as increase the tracking accuracy. The beam energy is concentrated towards the malicious target while moving farther using dynamic ranging. In addition, triangulation using two references dynamically allows changing references rather than the initialization of triangulation using three references. Thus, PL&T accuracy is significantly increased for multiple targets tracking even under multipath fading as compared to the existing PL&T using Omni-directional antennas.

The strictly friend reference nodes participate in PL&T, as soon as malicious nodes are detected, by developing dynamic tracking zone over the malicious nodes using the ratio of the distance between reference nodes and the average distance of target from reference nodes. The coverage of tracking zone is focused by sectored beams of two reference nodes until it goes outside the zone and later the zone is updated by using available sectored beams of reference nodes depending upon the target's direction of motion. The zone forming is rapidly iterated for multiple target's PL&T by the spatial reuse of multi-sectored adaptive beam forming.

## 2. Related Work

Movement Tracking System (MTS) is a critical battlefield navigator for combat service support by connecting mobile radio system mounted on tactical wheeled vehicle to control station at headquarters to provide PL&T information and route information about the mission to other soldier's radios. It consists of digital national geospatial-intelligence agency maps, global positioning system (GPS) location data, and L-band satellite two-way text messaging. MTS+ is advanced systems which has an embedded military GPS card and embedded radio frequency identification (RFID) interrogator within the new L-band satellite transceiver. MTS–II software enhancements provide better location tracking, flexible messaging and better command and control. Similarly, Blue Force Tracking (BFT) is the GPS-enabled system that provides military commanders with location information about friendly and hostile military forces on the computer's terrain-map display. The above mentioned movement's location tracking systems are governed by the situation awareness room to provide PL&T information in the graphical display and command control. However, these are not capable to detect malicious or hostile radios by radio detection and immediately multicast towards other friendly radios for tracking and alert from hostile threat in the distributed networks. In addition, there is risk as many countries have capability to jam and destroy satellite systems. GPS also does not work near or inside the building or metallic structure. This can be addressed by the integrated zone finding and triangulation over malicious or hostile radio deploying two friendly radios equipped with smart antennas and location is mapped [1-5]. The malicious and friendly radios are detected based on the encrypted anonymous private location assets [7-8].

Position Location and Tracking deploy different method for range measurement such as Angle of Arrival (AoA), Received Signal Strength (RSS), Time of Arrival (ToA), Time Difference of Arrival (TDoA) [1-5], [7].The location prediction based on these ranging method using one or two radios with directional antennas for estimation of the location of the target radio claims that it achieves better tracking accuracy over the triangulation based on using omni-directional antennas [1-3]. However, this technique has drawback of the lack of zone finding and increased accumulated errors in the computation of distance and the angle. In addition, the range measurements are not accurate in multipath faded channel due to uncorrelated signals, which is typically the case in the indoor tracking applications. These issues are addressed by the position location and tacking using the average of Time of Arrival (ToA) and Time of Departure (ToD) of the Internet Protocol (IP) based management packets in zone finding adaptive beam forming and triangulation.

In this paper, we propose robust strict friendly verification scheme for friendly and malicious nodes detection and the multiple malicious targets' PL&T scheme using MIMO antennas to concentrate the multiple beam forming over the predicted multiple tracking zones. The tracking zones are developed considering the current locations of two references as well as target, and the direction of movement within the battlefield mobility characteristics. The size of zone defines the adaptive beam width over zone to improve the performance under multi-path fading and improve the tracking accuracy by creating optimal multiple beams for triangulation in PL&T tracking based on ToD and ToA information. It also has the advantage of only using two references instead of three references for 2D PL&T and allows dynamic switching of reference nodes to maintain friendliness. Multiple nodes can be tracked by dynamically switching sectored beams between existing two or other friendly references.

## 3. Strict Friendly Verification (SFV) Scheme

Neighboring reference nodes are essential to verify them legitimately as strictly friends so that they can co-operate in the secured location tracking through multi-hop communications. The novel integrated key approach for packets encryption and decryption intending the strict friendliness verification deploy 256 bits length integrated key, which accomplish the minimal key lengths for symmetric ciphers to provide adequate commercial security. Initially, cluster head pre-initialize random number generator functions and symmetric keys. The novel integrated key is generated as 256 bits key K= ($K_1$ $K_2$ $K_3$) consisting 96 bits key ($K_1$), 64 bits key ($K_2$) and 96 bits key ($K_3$) to encrypt the packet by cluster head (CH). $K_1$ is pseudo random number using initial seed (i-1) in the $RNG_1$ function, $K_2$ is 64 bits secret symmetric key or identity $ID_{Tx}$ of a node and $K_3$ is generated by the encryption of the first packet with the random number generated from $RNG_2$ function using initial seed (n-1). Furthermore, the initial seed (i-1) for $K_{1,}$ is the location information (distance and direction) between CH and receiver whereas the initial seed (n-1) for $K_{2,}$ is the RTT of preamble packet between them. Then, first ensemble packet is encrypted using the first key K generated from the integration of the location information, RTT and symmetric key or ID. Similarly, the second key K´= ($K'_1$ $\|K'_2\|$ $K'_3$) is the integration of $K'_1$, generated from $RNG_1$ function using first halve of key K as seed i, $K'_2$ same as $K_2$ (symmetric key) and $K'_{3,}$ generated from $RNG_2$ function using second halve of key K as seed n. Then, the second ensemble packet is encrypted using second key K´ and this encryption procedure is iterated for next packets. The cipher packets $C_j$ for $j^{th}$ number of plain ensemble packets are computed using key $k_j$ in encryption as shown in Figure 1 using the equations 1 - 5.

$$K_{i1} = RNG_1(seed_i) \quad (1)$$

$$K_{i2} = ID_{Tx} \quad (2)$$

$$K_{i3} = \{ RNG_2(seed_n) \oplus P_{i1} \} \quad (3)$$

$$K_i = \{K_{i1} \| K_{i2} \| K_{i3}\} \quad (4)$$

$$C_i = P_i \oplus K_i \quad (5)$$

At the receiver end, the first packet is decrypted by using key K= ($K_1 \| K_2 \| K_3$) where, $K_1$ is the 96 bits key generated using initial seed (i-1) which is the location information between transmitter and receiver in $RNG_1$, $K_2$ is the 64 bits symmetric key of node and $K_3$ is 96 bits key generated by the decryption of the first encrypted packets with the random number generated from $RNG_2$ function using the initial seed (n-1) which is the RTT between transmitter and receiver. Regarding second packet, the decryption is done by changing $K'_1$ and $K'_3$ generated by RNG functions with new seeds achieved after halving the previous key and this decryption procedure is iterated for next packets. The plain ensemble packets $P_j$ for $j^{th}$ number of cipher packets are computed using keys $k_j$ in decryption as shown in Figure 2 using the following equations 6 - 10.

$$K_{j1} = RNG_1(seed_i) \quad (6)$$

$$K_{j2} = ID_{Rx} \quad (7)$$

$$K_{j3} = \{ RNG_2(seed_n) \oplus C_{j1} \} \quad (8)$$

$$K_j = \{K_{j1} \| K_{j2} \| K_{j3}\} \quad (9)$$

$$P_j = \{D_{kj}(C_j)\} = C_j \oplus K_j \quad (10)$$

The receiver is declared as a strict friendly neighbor node, when each encrypted packets from CH are successfully decrypted by the receiver and validated by CH which can be used for direct or multi hop communication. If not, the node is deemed as a malicious node and is not authorized for payload data communications. This process of strict friendliness verification between neighbors is required each time for secure path connectivity for information exchange.

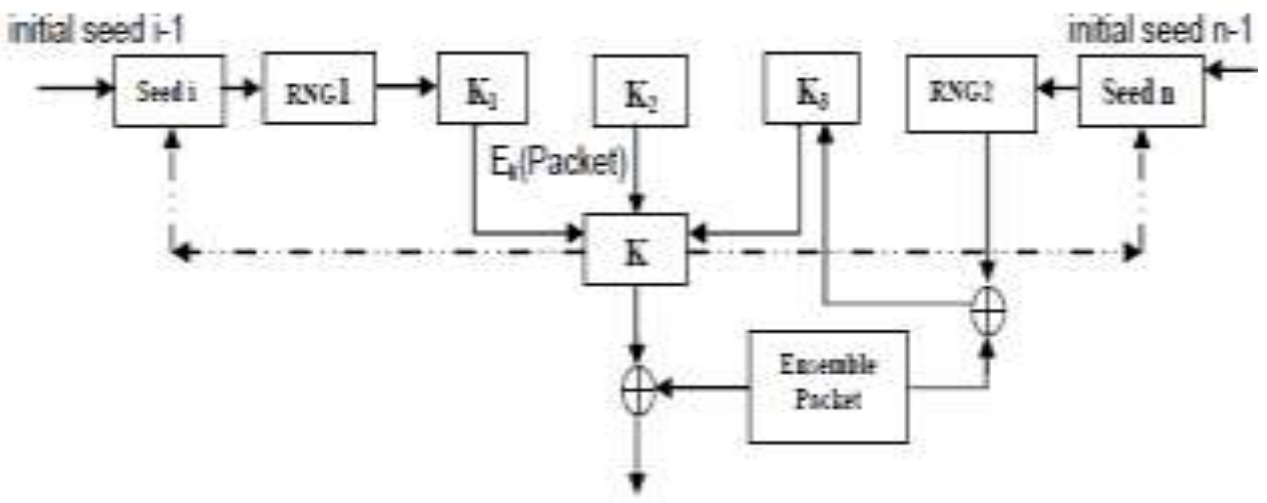

Figure 1: Packet Encryption for Strict Friendly Verification

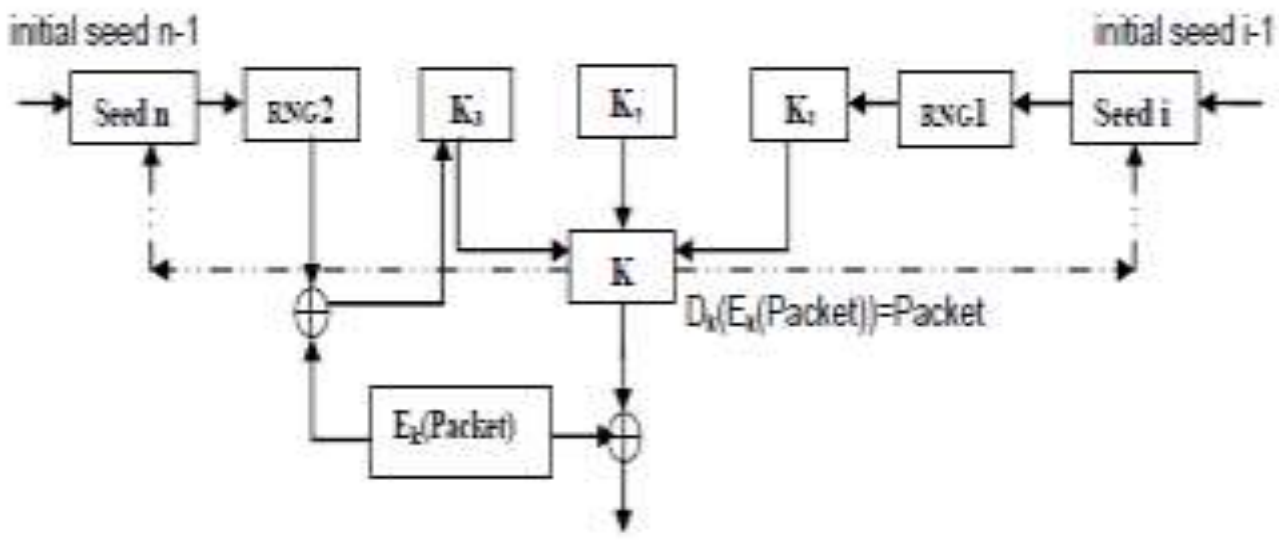

Figure 2: Packet Decryption for Strict Friendly Verification

Regarding the symmetric cipher, higher keys length protect against brute force attacks. Increasing each bit in the key increases twice the number of possible keys and yields two times more search for the brute force attack. With 256-bits key, the complexity analysis of algorithm needs O $(2^{256})$ = 1.15792089 × $10^{77}$ runs for the brutal force search. On average, a brute force attack must check half of the total runs, performing $2^{255}$ encryptions, to find the key. This 256 bits key length is enough for the symmetric ciphers to provide plentiful commercial security.

The detection rate is the probability of detecting a malicious or foe node in MANET cluster which evaluates the detection performance of strict friendliness verification scheme. When a suspicious transmitting node sends request to the receiver node, such that it could persuade as a friendly node and avoid being detected during the strict friendly verification process, with the probability of replayed by wormholes ($p_{wh}$), probability of node's key replay ($p_i$) and the probability of locally RTT replayed ($p_r$) by neighbors. Then, the probability of detection of replayed by wormholes is ($1-p_{wh}$), the probability of detection of node's key replay is ($1-p_i$) and the probability of detection of locally RTT replayed ($1-p_r$) in the strict friendliness verification. The probability of suspicious node detection by a friendly node is computed as:

$$P = (1-p_{wh})*(1-p_i)*(1-p_r) \quad (11)$$

When each detecting node having n detection keys, the detection rate or probability ($P_{dr}$) of a suspicious node being detected by a kind friendly detecting node can be computed as:

$$P_{dr} = 1-(1-p^n) = 1-\{1-(1-p_{wh})*(1-p_i)*(1-p_r)\}^n \quad (12)$$

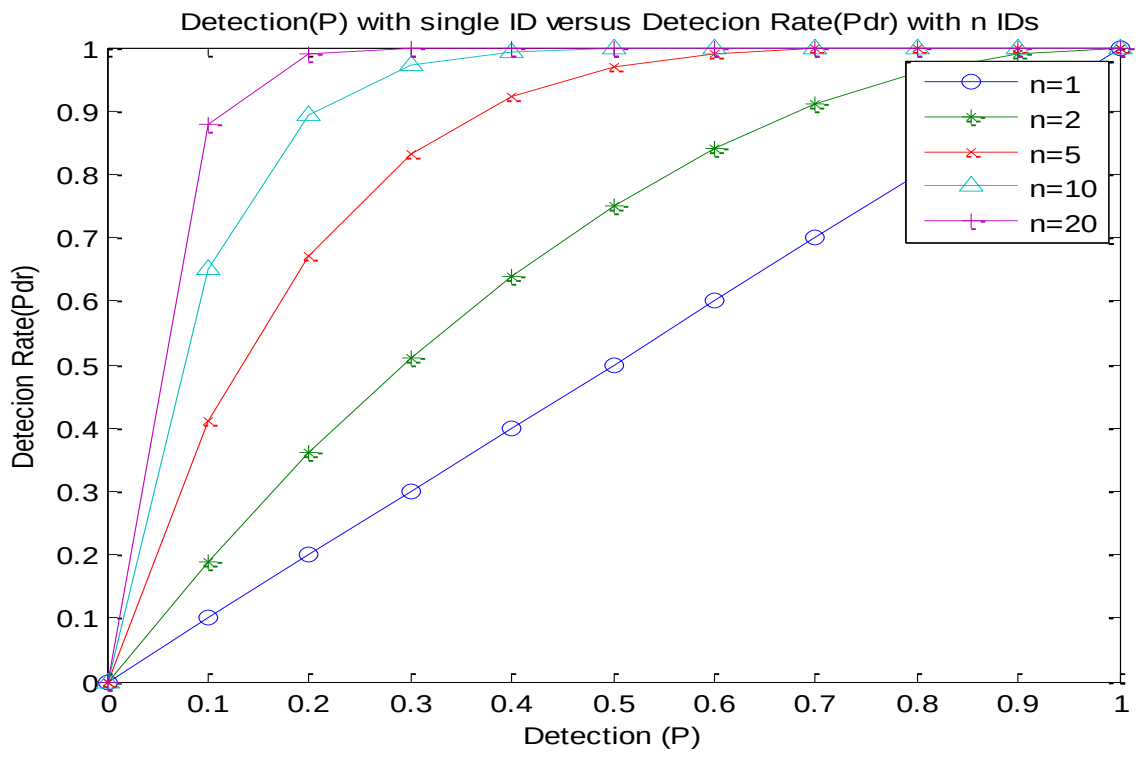

Figure 3: Detection Rate Analysis

This implies that unless the detection probability increases, the detection rate cannot be increased. A kind friendly detecting node can significantly increase the detection rate using higher number of symmetric keys. Figure 3 shows that the transmitter detection rate increases with the higher number of Symmetric keys changing at receiver because it increases the robustness against replay attacks.

## 4. Multiple Target's PL&T Algorithm

- Cluster head deploy SFV based authentication for each node by the exchange of ToD and ToA information to compute the Euclidean distance as well as RTT and symmetric keys a explained in section 4. If the node is not authenticated then it is malicious target node, otherwise strict friendly node.

- When the cluster head detects the malicious target $B_i$ then it assigns any two friendly (authenticated) nodes as references nodes $A_j$ and $C_k$. These are friendly till they participate in PL&T operation. These are further re-used as reference only when these can cover other malicious targets within threshold radius by multiple beamforming simultaneously. The reference nodes $A_j$ and $C_k$ send RTS packet using sectored antenna scanning for the target node $B_i$. Once, these nodes get CTS from the target $B_i$ then it is second reference node $C_k$ and retrieves management packets from target $B_i$ for ToA, ToD, DoA and data tracking. Otherwise, other friendly nodes are assigned by cluster head if target moves away from the range of the previous reference nodes.

- When both reference nodes $A_j$ and $C_k$ localize the fresh position of the target node $B_i$, then the tracking zone is formulated for Rapid PL&T whose radius $r_i$ is the ratio of the distance between reference nodes $d_{jk}$ and the average distance $d_{avg=}$ 0.5*($d_{ji}$+$d_{ki}$) from reference nodes to the target node $B_i$. This tactics enable that the coverage of tracking zone is increased at the nearest location of target $B_i$ from reference nodes $A_j$ and $C_k$ where the beam width is wider. Similarly, the coverage of tracking zone is decreased at the furthest location of target $B_i$ from reference nodes where the beam width is narrower.

- Once the tracking zone is formed then the reference nodes $A_j$ and $C_k$ focus their sectored beam over the tracking zone of the target $B_i$ by dynamic ranging to optimize the tracking dynamics and performance. Furthermore, the tracking zone can be tangentially shifted and changed as per target's directional motion. In addition, reference nodes and their sectored beam can be changed and switched for adjusting the tracking zone.

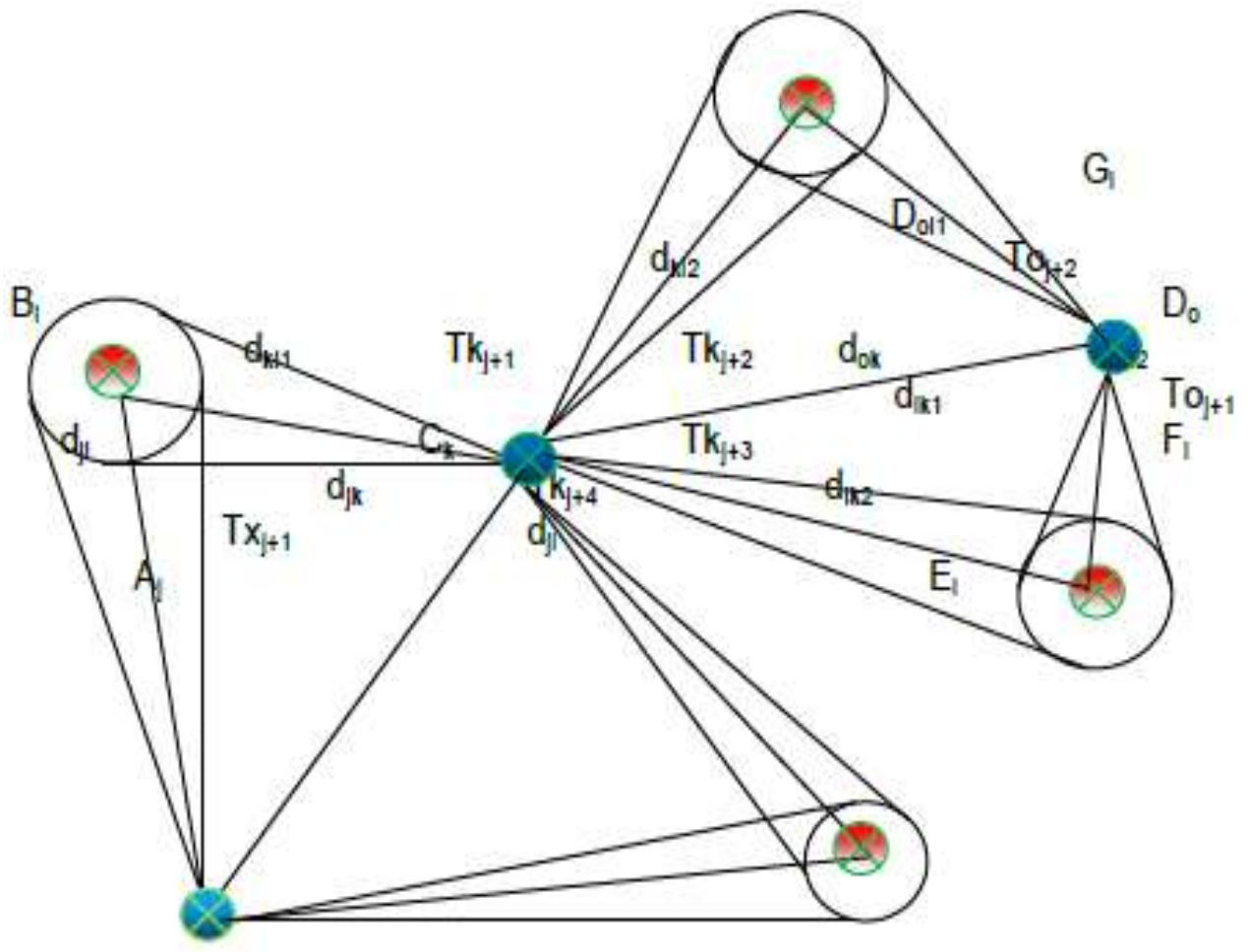

Figure 4: Multiple Targets PL&T using Four Sectored Antenna

All above steps are repeated by other sectored antennas { $Tx_j$ ….$Tx_{j+n-1}$ , $Tx_k$ ….$Tx_{k+n-1}$} of both reference nodes to form different tracking zones over different neighboring targets simultaneously using the spatial reuse of four sectored antennas.

Mathematically, the set of positions $P^k_{j,i}$ of different i targets by a node deploying k different sectored antenna beams along with j reference nodes are computed as below. The targets are assumed to be moving at velocity $v^k_i$ in the direction $\theta^{\check{}k}_{j,i}$ with Euclidean distance $D^k_{j,i}$ at time $\Delta t^k_{j,i}$ respectively as shown in Figure 4.

$$\bigcup_{i,j,k=1}^{i,k=4;j\leq 4} P^k_{i,j} \tag{13}$$

$$\bigcup_{i,j,k=1}^{i,k=4;j\leq 4} v^k_i \; (\theta^{\check{}k}_{j,i})\Delta t^k_{j,i} \tag{14}$$

$$\bigcup_{i,j,k=1}^{i,k=4;j\leq 4} v^k_i *(\cos(\theta^k_{j,i})x^{\check{}}_{j,i} + (\sin\theta^k_{j,i})y^{\check{}}_{j,i})*\Delta t^k_{j,i} \quad (14) \tag{15}$$

The efficiency of Rapid PL&T is evaluated over 100 different random experiments of PL&T observations by changing the position of tracking nodes and tracking zone regarding the target. The result is found that the average efficiency of Rapid PL&T is 96% as shown in Figure 5. Therefore, the proposed Rapid PL&T is highly efficient and faster PL&T algorithm.

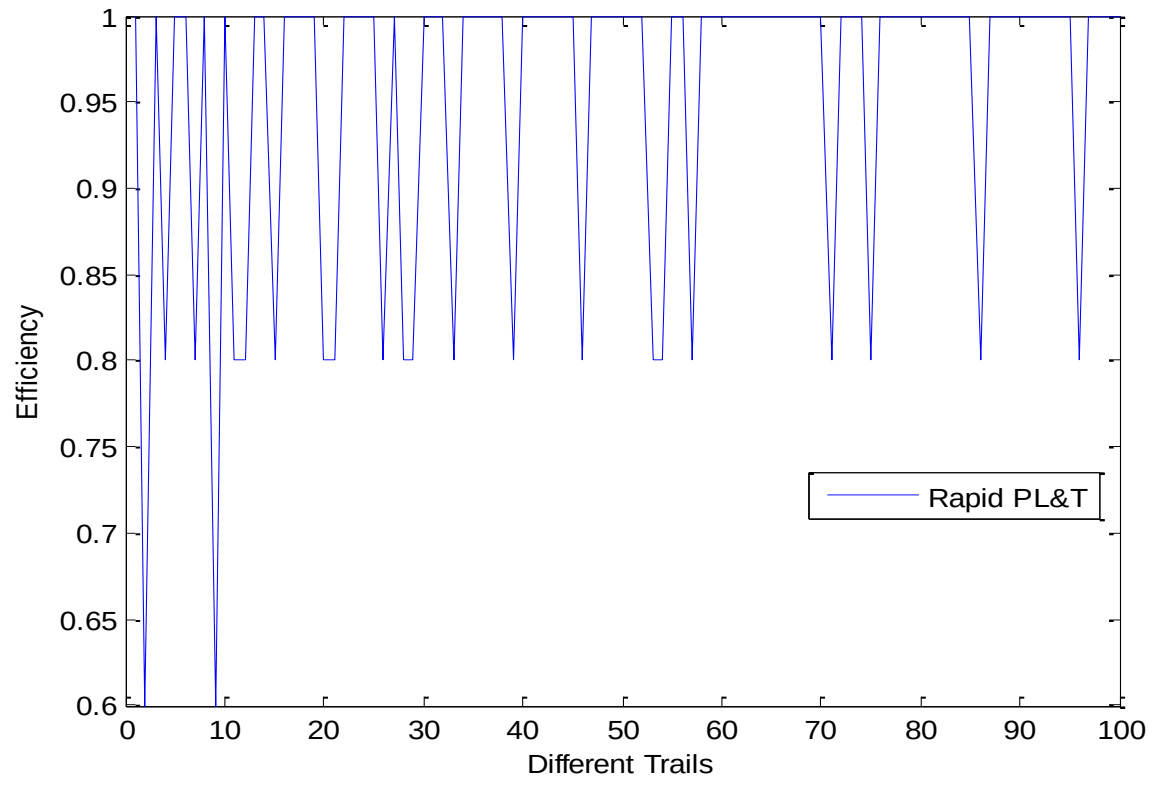

Figure 5: Rapid PL&T using Sectored Array Antennas

## 5. Simulation & Performance Evaluation

Multiple target's PL&T is deployed in cluster size of 400 X 400 sq. m. deploying 60 nodes with 250 m range in each cluster with similar mobility characteristics deploying iteratively Rapid PL&T at each sector of multiple sectored antennas with the friendly nodes. This enables that each reference node can localize and track four targets simultaneously at its different sectored beams with other friendly reference nodes. Simulation shows that iterative deployment of Rapid PL&T at four different sectored antennas of a reference node provides average PL&T efficiency of 96%, 85%, 69% and 60% for four different targets over 100 different random experiments taken at the interval of 5 seconds in 500 seconds by changing the position of targets and the corresponding tracking zone as shown in Figure 6.

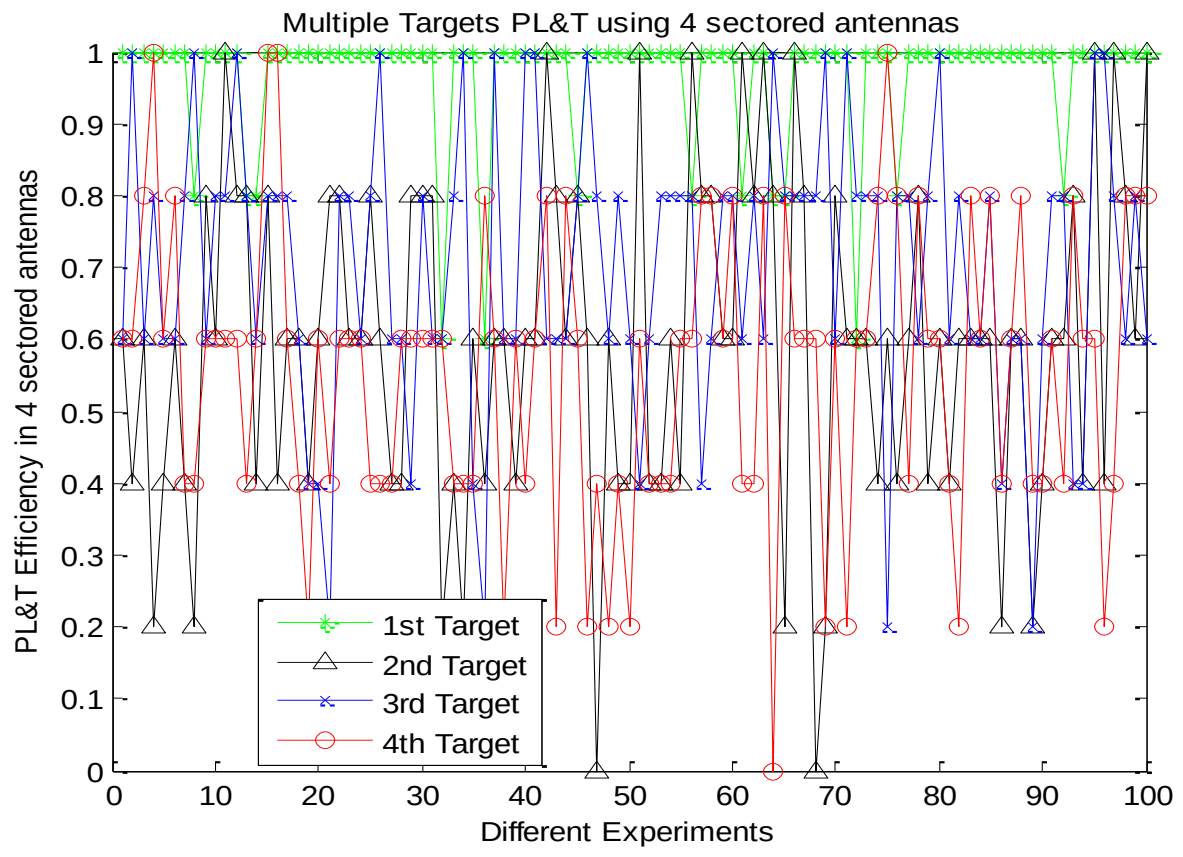

Figure 6: Multiple PL&T using Four Sectored Antennas

Multiple Targets PL&T is deployed with 4 multiple sectored antennas and switching overhead is simulated regarding velocity of target nodes in 50 different experiments in Figure 7-8. Simulation results show that the switching overhead is higher at higher velocity of the targets because the reference nodes are rapidly switched in out of range scenario. The overhead is higher even at speed of 20 m/s because one node is localizing and tracking four random target nodes with other four reference making four different pairs by switching beams as per availability and requirement. This concurs that the low relative speed bound of a single node using four consecutive adaptive beams increased the switching overhead in Multiple Targets PL&T.

It is observed that the first target is easily detected and tracked with another reference node switching the sectored beam and later reference node is switched at 11th and 25th experiments. Similarly, second target is detected and tracked with another reference node switching remaining three sectors and later reference node is switched at 18th experiment. Third target is also detected and tracked with other reference node and switching desired sector beam from two sectors and later reference node is switched at 43th experiment. Fourth target took long overhead at the initial experiment to scan the desired reference node to detect and track using the last sector and later changed reference node at 35th experiment. The beam switching is deployed as soon as sectored beams are available as per requirement of reference nodes and targets.

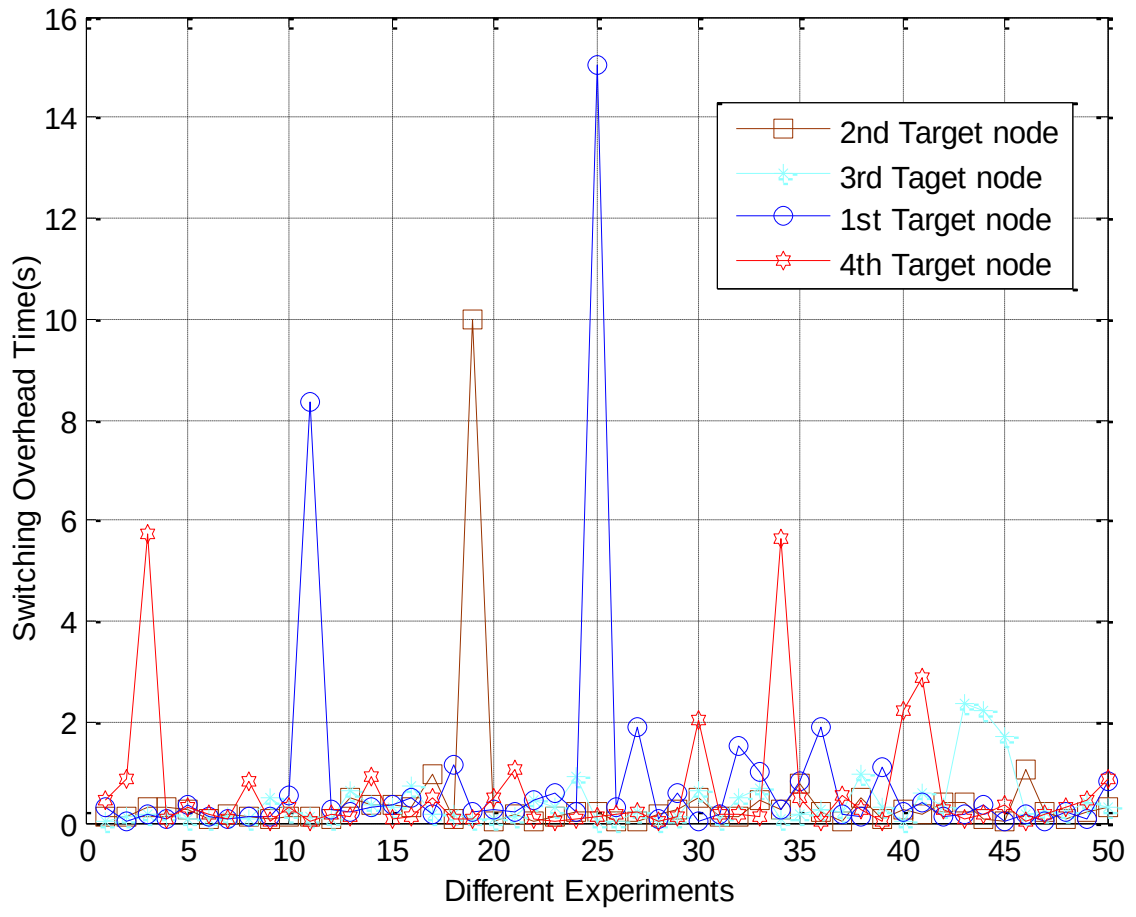

Figure 7: Switching Overhead of Multiple Targets PL&T

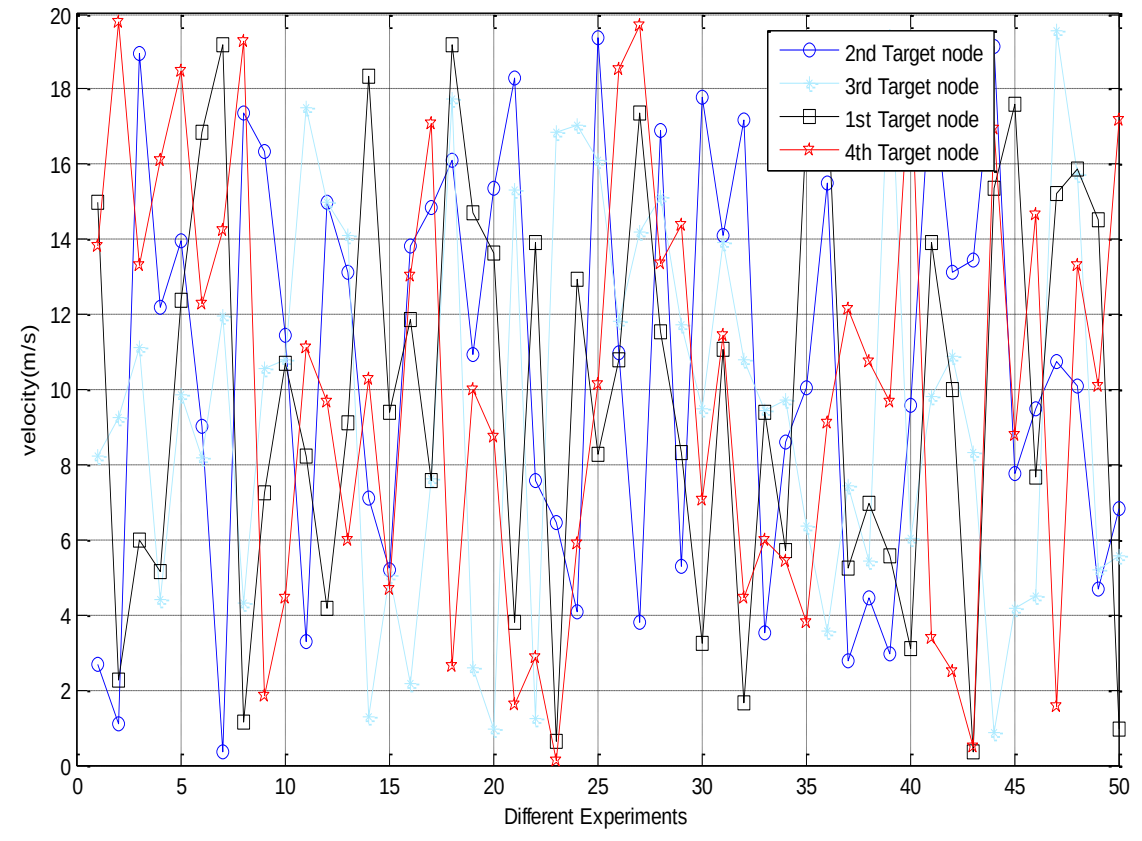

Figure 8: Velocity of Multiple Targets PL&T

Multiple Target's Trajectory and their PL&T estimation are illustrated in Figure 9 for four different target nodes moving along parallel paths. The average errors in PL&T estimation of four different targets trajectory are found to be 1.73 m, 2.13 m, 2.22 m and 2.33 m respectively. In addition, the received signal energy by a receiving target node is analyzed by varying different number of sectored beams at transmitting node as illustrated in Figure 10. The simulation results show that the received energy decreases with increasing number of the sectored beams because of the inter signal interference between sectored beams.

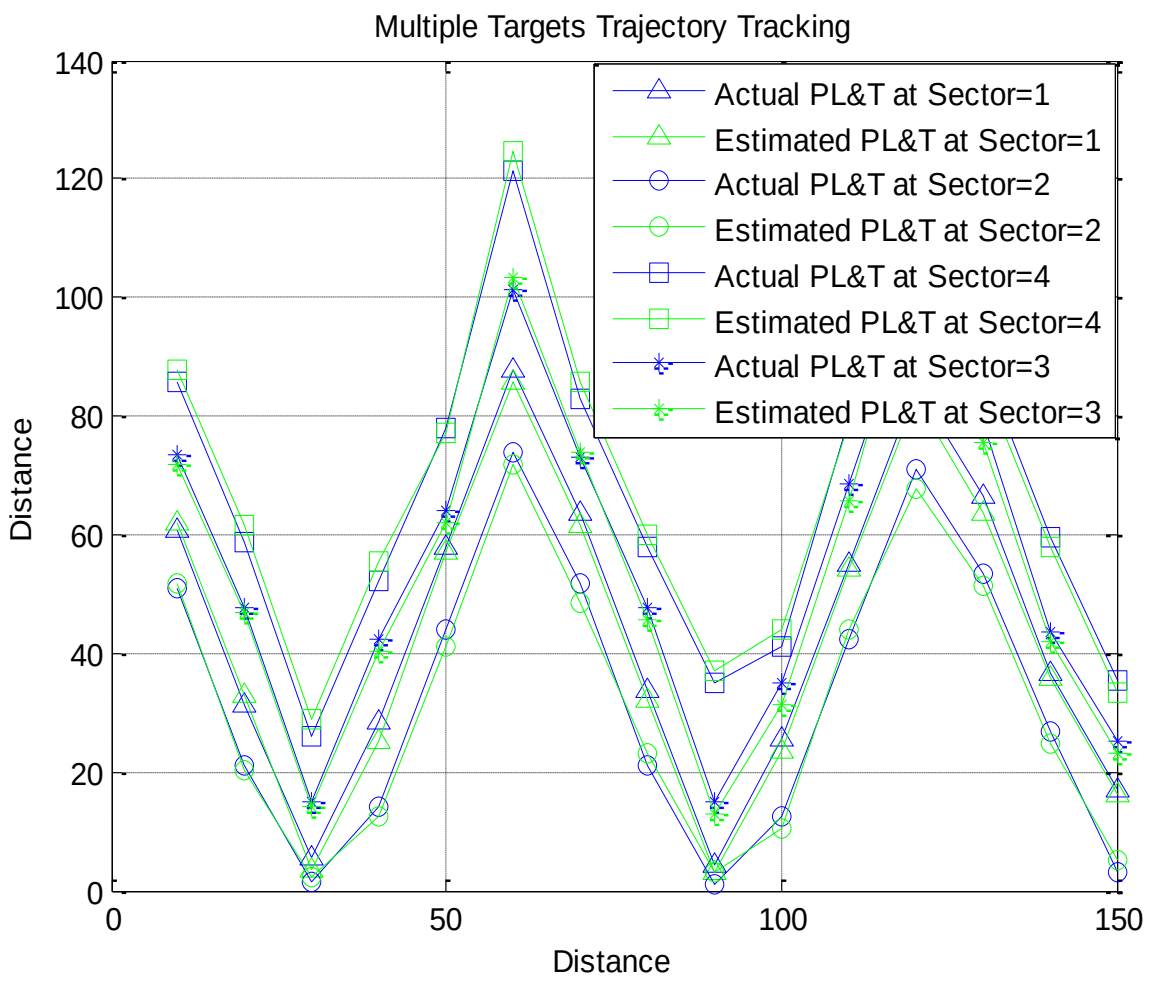

Figure 9: Multiple Targets Trajectory PL&T

Friendliness between reference nodes is essential for developing the narrow zones, adaptive beam forming and tracking of the multiple malicious nodes using 4 sectored antennas. Figure 11 depicts the maintenance of friendliness by 4 sectored beams of a reference node with another 4 different reference nodes. The first three friendly nodes are authenticated using different beams from sector-1,sector-2, sector-3 easily but the fourth node need at least 20 seconds to scan and authenticate new friendly nodes as three friendly nodes already in use by other sectored beams. Furthermore, when the existing reference node does not maintain friendliness for PL&T then another friendly node is authenticated and use for PL&T. The true friendliness is maintained only when the node is authenticated based on SFV. The friendliness authentication overhead depends upon the time period between two latest failures in maintaining friendliness as illustrated in Figure 11. The scanning and authentication overhead is at least 20 seconds for a single friendliness failure after some time interval and at least 30 seconds for any two consecutive friendliness failures.

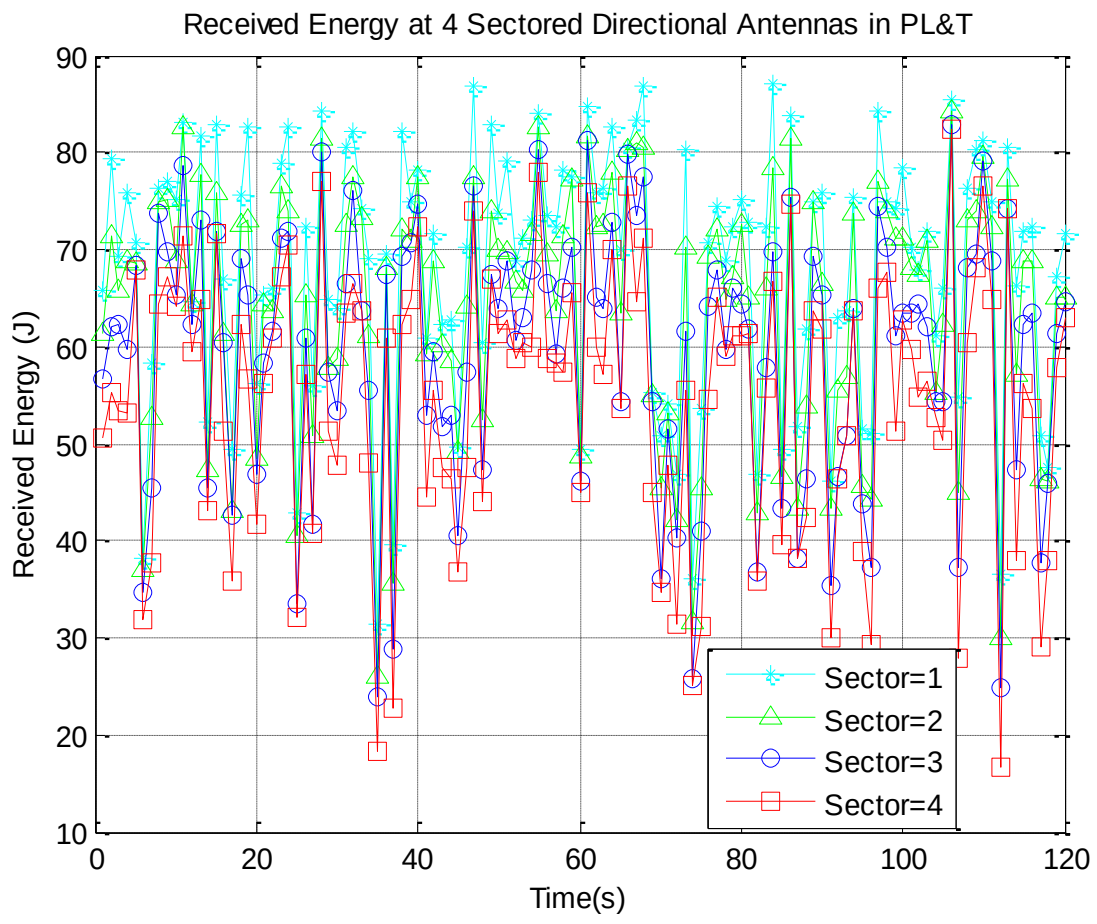

Figure 10: Received Signal Energy at Different Sectored Beams

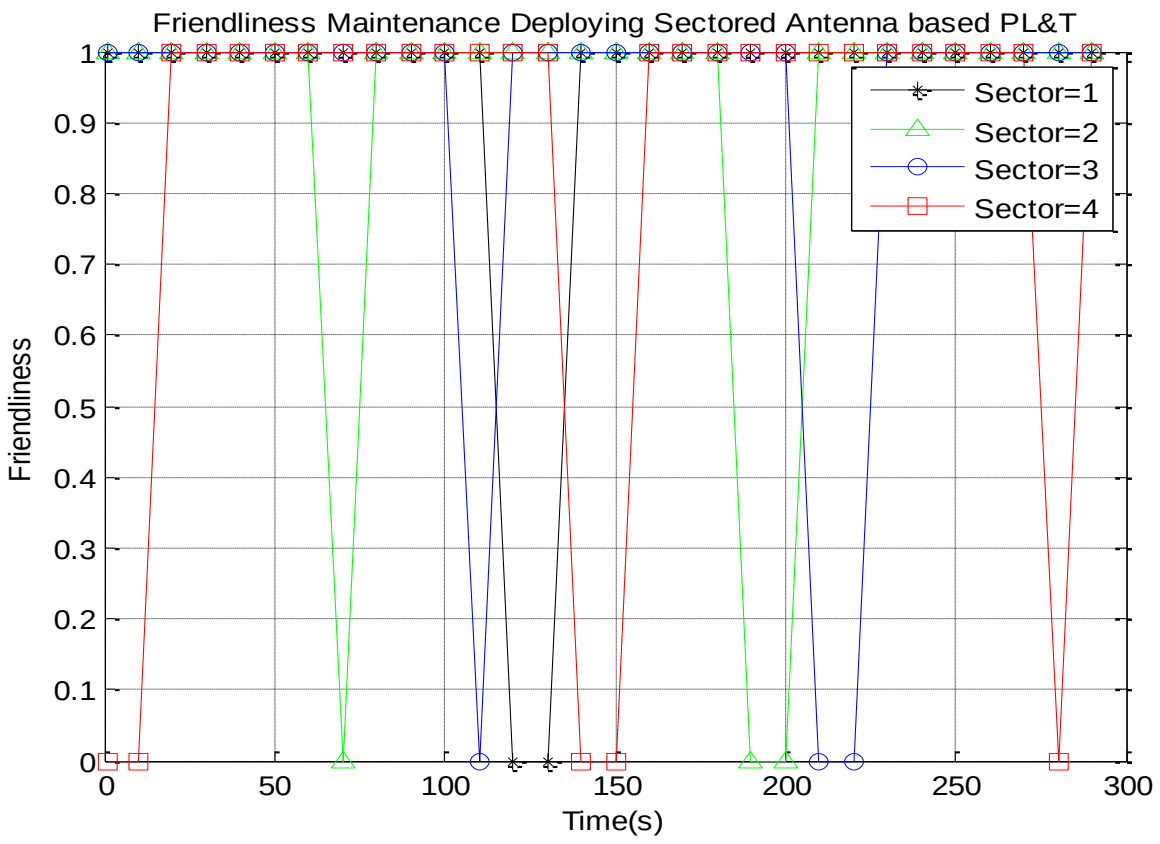

Figure 11: Maintaining Friendly Reference Nodes during PL&T

.

## 6. Conclusion

Multiple targets' PL&T integrated with strict friendly verification significantly increase the PL&T detection rate by increasing the number of symmetric keys as well as rapid zone finding for multiple malicious nodes. It is found that the low relative speed bound of any single node using four consecutive adaptive beams increased the switching overhead in multiple targets' PL&T. The beam switching is deployed as soon as sectored beams are available as per requirement of reference nodes and targets. In addition, the received energy decreases with increasing number of the sectored beams because of the inter signal interference between sectored beams. Furthermore, strict friendly authentication overhead depends upon the time period between two latest failures in maintaining friendliness. Future research will concentrate on the secured location aware multimedia streaming with QoS optimization.

## ACKNOWLEDGMENT

This research work is supported in part by the U.S. ARO under Cooperative Agreement W911NF-04-2- 0054 and the National Science Foundation NSF 0931679. The views and conclusions contained in this document are those of the authors and should not be interpreted as representing the official policies, either expressed or implied, of the Army Research Office or the National Science Foundation or the U. S. Government.

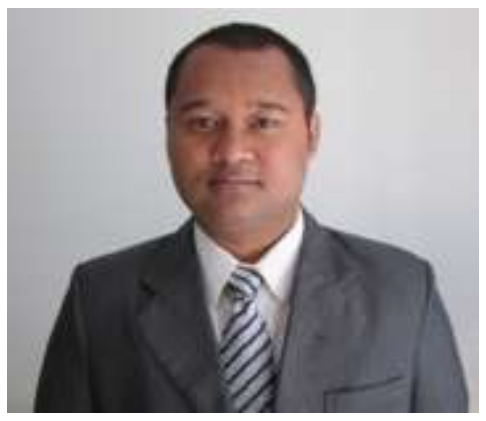

**Dr. Niraj Shakhakarmi** worked as a Doctoral Researcher since 2009-2011 in the US-Army Research Office (ARO) Center for Digital Battlefield Communications (CeBCom) Research, Department of Electrical and Computer Engineering, Prairie View A&M University. He received his B.E. degree in Computer Engineering in 2005 and M.Sc. in Information and Communication Engineering in 2007. His research interests are in the areas of Wireless Communications and Networks Security, Secured Position Location & Tracking (PL&T) of Malicious Radios, Cognitive Radio Networks, WCDMA/HSPA/LTE/WRAN Next Generations Wireless Networks, Satellite Networks and Digital Signal Processing, Wavelets Applications and Image/Colour Technology. He is a member of IEEE Communications Society, ISOC, IAENG and attended AMIE conference. He has published several WSEAS journals and IJCSI journals, along with WTS, Elsevier and ICSST conference papers. His several journals and conference papers are under review in IEEE journals. He is serving as reviewer for WSEAS, SAE, WTS and editorial board member for IJEECE, WASET, JCS and IJCN journals.

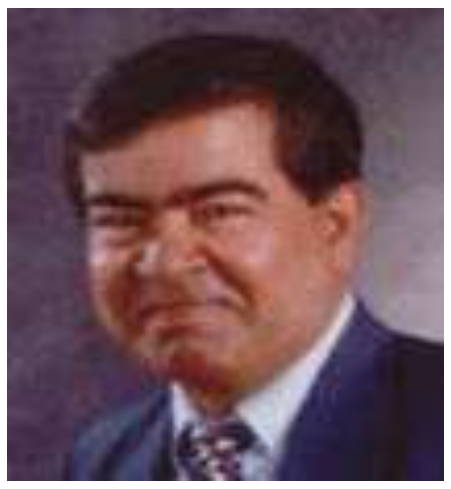

**Prof. Dhadesugoor R. Vaman** is Texas A & M University Board of Regents and Texas Instrument Endowed Chair Professor and Founding Director of ARO Center for Battlefield Communications (CeBCom) Research, ECE Department, Prairie View A&M University (PVAMU). He has more than 38 years of research experience in telecommunications and networking area. Currently, he has been working on the control based mobile ad hoc and sensor networks with emphasis on achieving bandwidth efficiency using KV transform coding; integrated power control, scheduling and routing in cluster based network architecture; QoS assurance for multi-service applications; and efficient network management.

Prior to joining PVAMU, Dr. Vaman was the CEO of Megaxess (now restructured as MXC) which developed a business ISP product to offer differentiated QoS assured multi-services with dynamic bandwidth management and successfully deployed in several ISPs. Prior to being a CEO, Dr. Vaman was a Professor of EECS and founding Director of Advanced Telecommunications Institute, Stevens Institute of Technology (1984 1998); Member, Technology Staff in COMSAT (Currently Lockheed Martin) Laboratories (1981- 84) and Network Analysis Corporation (CONTEL) (1979-81); Research Associate in Communications Laboratory, The City College of New York (1974- 79); and Systems Engineer in Space Applications Center (Indian Space Research Organization) (1971- 1974). He was also the Chairman of IEEE 802.9 ISLAN Standards Committee and made numerous technical contributions and produced 4 standards. Dr. Vaman has published over 200 papers in journals and conferences; widely lectured nationally and internationally; has been a key note speaker in many IEEE and other conferences, and industry forums. He has received numerous awards and patents, and many of his innovations have been successfully transferred to industry for developing commercial products.